\input harvmac.tex

\input epsf.tex

\def\figin{\epsfcheck\figin}\def\figins{\epsfcheck\figins}
\def\epsfcheck{\ifx\epsfbox\UnDeFiNeD
\message{(NO epsf.tex, FIGURES WILL BE IGNORED)}
\gdef\figin##1{\vskip2in}\gdef\figins##1{\hskip.5in}
\else\message{(FIGURES WILL BE INCLUDED)}%
\gdef\figin##1{##1}\gdef\figins##1{##1}\fi}
\def\DefWarn#1{}
\def\figinsert{\goodbreak\midinsert}
\def\ifig#1#2#3{\DefWarn#1\xdef#1{fig.~\the\figno}
\writedef{#1\leftbracket fig.\noexpand~\the\figno}%
\figinsert\figin{\centerline{#3}}\medskip\centerline{\vbox{\baselineskip12pt
\advance\hsize by -1truein\noindent\footnotefont{\bf Fig.~\the\figno:} #2}}
\bigskip\endinsert\global\advance\figno by1}

\def\ov{\over }
\def\a{\alpha }
\def\s{\sigma }



\lref\ArutyunovUJ{
G.~Arutyunov, S.~Frolov, J.~Russo and A.~A.~Tseytlin,
``Spinning strings in AdS(5) x S**5 and integrable systems,''
Nucl.\ Phys.\ B {\bf 671}, 3 (2003)
[arXiv:hep-th/0307191].}

\lref\ART{G.~Arutyunov, J.~Russo and A.~A.~Tseytlin,
``Spinning strings in AdS(5) x S**5: New integrable system relations,''
Phys.\ Rev.\ D {\bf 69}, 086009 (2004)
[arXiv:hep-th/0311004].}

\lref\AlishahihaVI{
M.~Alishahiha, A.~E.~Mosaffa and H.~Yavartanoo,
``Multi-spin string solutions in AdS black hole and confining backgrounds,''
arXiv:hep-th/0402007.
}

\lref\GKP{
S.~S.~Gubser, I.~R.~Klebanov and A.~M.~Polyakov,
``A semi-classical limit of the gauge/string correspondence,''
Nucl.\ Phys.\ B {\bf 636}, 99 (2002)
[arXiv:hep-th/0204051].
}

\lref\armoni{
A.~Armoni,
``Witten-Veneziano from Green-Schwarz,''
JHEP {\bf 0406}, 019 (2004)
[arXiv:hep-th/0404248].}

\lref\chir{
J.~Babington, J.~Erdmenger, N.~J.~Evans, Z.~Guralnik and I.~Kirsch,
``Chiral symmetry breaking and pions in non-supersymmetric gauge /  gravity
duals,''
Phys.\ Rev.\ D {\bf 69}, 066007 (2004)
[arXiv:hep-th/0306018].}

\lref\MaldacenaRE{
J.~M.~Maldacena,
``The large N limit of superconformal field theories and supergravity,''
Adv.\ Theor.\ Math.\ Phys.\  {\bf 2}, 231 (1998)
[Int.\ J.\ Theor.\ Phys.\  {\bf 38}, 1113 (1999)]
[arXiv:hep-th/9711200].
}

\lref\PeetWN{
A.~W.~Peet and J.~Polchinski,
``UV/IR relations in AdS dynamics,''
Phys.\ Rev.\ D {\bf 59}, 065011 (1999)
[arXiv:hep-th/9809022].
}

\lref\KlebanovHB{
I.~R.~Klebanov and M.~J.~Strassler,
``Supergravity and a confining gauge theory: Duality cascades and
chiSB-resolution of naked singularities,''
JHEP {\bf 0008}, 052 (2000)
[arXiv:hep-th/0007191].
}

\lref\ItzhakiDD{
N.~Itzhaki, J.~M.~Maldacena, J.~Sonnenschein and S.~Yankielowicz,
``Supergravity and the large N limit of theories with sixteen  supercharges,''
Phys.\ Rev.\ D {\bf 58}, 046004 (1998)
[arXiv:hep-th/9802042].
}

\lref\SusskindDQ{
L.~Susskind and E.~Witten,
``The holographic bound in anti-de Sitter space,''
arXiv:hep-th/9805114.
}

\lref\MaldacenaIM{
J.~M.~Maldacena,
``Wilson loops in large N field theories,''
Phys.\ Rev.\ Lett.\  {\bf 80}, 4859 (1998)
[arXiv:hep-th/9803002].
}

\lref\ReyIK{
S.~J.~Rey and J.~T.~Yee,
``Macroscopic strings as heavy quarks in large N gauge theory and  anti-de
Eur.\ Phys.\ J.\ C {\bf 22}, 379 (2001)
[arXiv:hep-th/9803001].
}

\lref\BrandhuberER{
A.~Brandhuber, N.~Itzhaki, J.~Sonnenschein and S.~Yankielowicz,
``Wilson loops, confinement, and phase transitions in large N gauge  theories
from supergravity,''
JHEP {\bf 9806}, 001 (1998)
[arXiv:hep-th/9803263].
}

\lref\minahan{
J.~A.~Minahan,
``Circular semiclassical string solutions on AdS(5) x S**5,''
Nucl.\ Phys.\ B {\bf 648}, 203 (2003)
[arXiv:hep-th/0209047].}

\lref\mina{
J.~A.~Minahan,
``Glueball mass spectra and other issues for supergravity duals of {QCD}
models,''
JHEP {\bf 9901}, 020 (1999)
[arXiv:hep-th/9811156].}

\lref\frolov{S.~Frolov and A.~A.~Tseytlin,
``Rotating string solutions: AdS/CFT duality in non-supersymmetric  sectors,''
Phys.\ Lett.\ B {\bf 570}, 96 (2003)
[arXiv:hep-th/0306143].}

\lref\KruczenskiUQ{
M.~Kruczenski, D.~Mateos, R.~C.~Myers and D.~J.~Winters,
``Towards a holographic dual of large-N(c) QCD,''
arXiv:hep-th/0311270.
}

\lref\KarchSH{
A.~Karch and E.~Katz,
``Adding flavor to AdS/CFT,''
JHEP {\bf 0206}, 043 (2002)
[arXiv:hep-th/0205236].
}

\lref\KruczenskiBE{
M.~Kruczenski, D.~Mateos, R.~C.~Myers and D.~J.~Winters,
``Meson spectroscopy in AdS/CFT with flavour,''
JHEP {\bf 0307}, 049 (2003)
[arXiv:hep-th/0304032].
}

\lref\BachasXS{
C.~Bachas,
``Convexity Of The Quarkonium Potential,''
Phys.\ Rev.\ D {\bf 33}, 2723 (1986).
}

\lref\ChernodubBK{
M.~N.~Chernodub, F.~V.~Gubarev, M.~I.~Polikarpov and V.~I.~Zakharov,
``Confinement and short distance physics,''
Phys.\ Lett.\ B {\bf 475}, 303 (2000)
[arXiv:hep-ph/0003006].
}

\lref\Spins{N.~Beisert, J.~A.~Minahan, M.~Staudacher and K.~Zarembo,
``Stringing spins and spinning strings,''
JHEP {\bf 0309}, 010 (2003)
[arXiv:hep-th/0306139];
S.~Frolov and A.~A.~Tseytlin,
``Rotating string solutions: AdS/CFT duality in non-supersymmetric  sectors,''
Phys.\ Lett.\ B {\bf 570}, 96 (2003)
[arXiv:hep-th/0306143];
G.~Arutyunov, S.~Frolov, J.~Russo and A.~A.~Tseytlin,
``Spinning strings in AdS(5) x S**5 and integrable systems,''
Nucl.\ Phys.\ B {\bf 671}, 3 (2003)
[arXiv:hep-th/0307191];
M.~Kruczenski, A.~V.~Ryzhov and A.~A.~Tseytlin,
``Large spin limit of AdS(5) x S**5 string theory and low energy expansion of
ferromagnetic spin chains,''
arXiv:hep-th/0403120;
G.~Arutyunov and M.~Staudacher,
``Two-loop commuting charges and the string / gauge duality,''
arXiv:hep-th/0403077.}

\lref\BornIV{
K.~D.~Born, E.~Laermann, N.~Pirch, T.~F.~Walsh and P.~M.~Zerwas,
``Hadron Properties In Lattice QCD With Dynamical Fermions,''
Phys.\ Rev.\ D {\bf 40}, 1653 (1989).
}

\lref\witten{E.~Witten,
``Anti-de Sitter space, thermal phase transition, and confinement in  gauge
theories,''
Adv.\ Theor.\ Math.\ Phys.\  {\bf 2}, 505 (1998)
[arXiv:hep-th/9803131].}

\lref\gross{D.~J.~Gross and H.~Ooguri,
``Aspects of large N gauge theory dynamics as seen by string theory,''
Phys.\ Rev.\ D {\bf 58}, 106002 (1998)
[arXiv:hep-th/9805129].}

\lref\csaki{C.~Csaki, H.~Ooguri, Y.~Oz and J.~Terning,
``Glueball mass spectrum from supergravity,''
JHEP {\bf 9901}, 017 (1999)
[arXiv:hep-th/9806021].}

\lref\ooguri{H.~Ooguri, H.~Robins and J.~Tannenhauser,
``Glueballs and their Kaluza-Klein cousins,''
Phys.\ Lett.\ B {\bf 437}, 77 (1998)
[arXiv:hep-th/9806171].}

\lref\russo{J.~G.~Russo,
``New compactifications of supergravities and large N {QCD},''
Nucl.\ Phys.\ B {\bf 543}, 183 (1999)
[arXiv:hep-th/9808117].}

\lref\csakrus{C.~Csaki, Y.~Oz, J.~Russo and J.~Terning,
``Large N {QCD} from rotating branes,''
Phys.\ Rev.\ D {\bf 59}, 065012 (1999)
[arXiv:hep-th/9810186].}

\lref\sfet{J.~G.~Russo and K.~Sfetsos,
``Rotating D3 branes and {QCD} in three dimensions,''
Adv.\ Theor.\ Math.\ Phys.\  {\bf 3}, 131 (1999)
[arXiv:hep-th/9901056].
}

\lref\csakk{C.~Csaki, J.~Russo, K.~Sfetsos and J.~Terning,
``Supergravity models for 3+1 dimensional {QCD},''
Phys.\ Rev.\ D {\bf 60}, 044001 (1999)
[arXiv:hep-th/9902067].}

\lref\hashi{A.~Hashimoto and Y.~Oz,
``Aspects of {QCD} dynamics from string theory,''
Nucl.\ Phys.\ B {\bf 548}, 167 (1999)
[arXiv:hep-th/9809106].
}

\lref\barbon{J.~L.~F.~Barbon, C.~Hoyos, D.~Mateos and R.~C.~Myers,
``The holographic life of the eta',''
arXiv:hep-th/0404260.
}

\lref\MN{J.~M.~Pons and P.~Talavera,
``Semi-classical string solutions for N = 1 SYM,''
Nucl.\ Phys.\ B {\bf 665}, 129 (2003)
[arXiv:hep-th/0301178];
L.~A.~Pando Zayas, J.~Sonnenschein and D.~Vaman,
``Regge trajectories revisited in the gauge / string correspondence,''
Nucl.\ Phys.\ B {\bf 682}, 3 (2004)
[arXiv:hep-th/0311190];
F.~Bigazzi, A.~L.~Cotrone and L.~Martucci,
arXiv:hep-th/0403261.}

\lref\bando{M.~Bando, T.~Kugo, A.~Sugamoto and S.~Terunuma,
``Pentaquark baryons in string theory,''
arXiv:hep-ph/0405259.}

\lref\barb{
A.~Armoni, J.~L.~F.~Barbon and A.~C.~Petkou,
``Rotating strings in confining AdS/CFT backgrounds,''
JHEP {\bf 0210}, 069 (2002)
[arXiv:hep-th/0209224].}


{ \Title{\vbox{\baselineskip12pt
{\vbox{\baselineskip12pt
}} }}
 {\vbox{
{\centerline { Semiclassical string spectrum}}
 {\centerline {in a  string model dual to large $N$  QCD}}
}}}
\bigskip
\centerline{ J.~M.~Pons$^{1}$, J.G.~Russo$^{1,2}$ and P.~Talavera$^{3}$ }
\bigskip
\centerline{$^1$ Departament d'Estructura i Constituents de la Mat\`eria,}
\centerline{ Universitat de Barcelona,
Diagonal 647, E-08028 Barcelona, Spain }

\medskip

\centerline{$^2$ Instituci\' o Catalana de Recerca i Estudis
  Avan\c{c}ats (ICREA)}
\medskip

\centerline{$^3$ Departament de F{\'\i}sica i Enginyeria Nuclear,}
\centerline{ Universitat Polit\`ecnica de Catalunya,
Jordi Girona 1--3, E-08034 Barcelona, Spain }

\vskip .3in

\baselineskip12pt

We explore the string spectrum in the Witten QCD$_4$ model
by considering classical string configurations, thereby
obtaining energy formulas for
quantum states with large excitation quantum numbers
 representing
glueballs and Kaluza-Klein states.
In units of the string tension, 
 the energies of all states increase as the 't Hooft coupling $\lambda $ is 
decreased, except the energies of 
 glueballs
corresponding to strings lying on the horizon, which remain constant. 
We argue that 
some  string solutions can be extrapolated to
the small $\lambda  $ regime.
We also find the classical mechanics description of
supergravity glueballs in terms of
point-like string configurations oscillating in the radial direction,
and  reproduce the glueball energy formula previously obtained by
solving the equation for the dilaton fluctuation.

\vfill

\Date{June  2004}
\eject
\baselineskip14pt
\newsec{Introduction}

Using the AdS/CFT correspondence \MaldacenaRE ,
Witten  proposed a string model representing a holographic dual to
 pure $SU(N)$ $3+1$ Yang-Mills theory \witten .
The idea is to start with $N$ D4 branes compactified on a circle of
radius
$r_0$, and
to impose anti-periodic boundary conditions for the fermions.
Then fermions get masses of order $r_0^{-1}$ and
scalar particles get masses at one loop. The resulting
low energy theory on the D4 brane is then pure $SU(N)$ Yang-Mills theory
in 3+1 dimensions, describing the dynamics of massless gluons.
The corresponding supergravity background can be constructed in terms
of the near-horizon limit of the euclidean non-extremal D4 brane.
This prescribes the correct
anti-periodic boundary conditions for the fermions.

The string spectrum in this supergravity background contains
particles with Kaluza-Klein  $S^4$ quantum numbers. These
should decouple, since they have no counterpart in QCD, which
does not have $SO(5)$ global symmetry.
The masses of these particles have been computed
in the supergravity sector \refs{\csaki,\ooguri ,\mina }, and, in 
the supergravity
approximation, they are basically
of the same order of glueball masses.
In addition, there are  Kaluza-Klein particles with $S^1$
quantum numbers which should also decouple, in order for the
string model to describe 3+1 dimensional  rather than 4+1 dimensional
Yang-Mills theory. They also have masses of the same order as glueball
masses. Such states can be decoupled by adding
rotation parameters to the supergravity background, which  breaks
the non abelian $SO(5)$ global symmetries to the Cartan subgroup
$U(1)\times U(1)$ \refs{\russo, \csakrus ,\sfet ,\csakk }.
But even in this case, the $S^4$ Kaluza-Klein modes
do not decouple within the supergravity approximation.

Nevertheless, the model reproduces in a quite remarkable way
several qualitative and quantitative features of QCD.
It describes the correct $N$ dependence of physical quantities
like e.g. a hadron spectrum independent of $N$
\witten\ and a gluon condensate proportional to $N^2$ \hashi .
It also exhibits confinement in the form of
an area law for Wilson loops 
\witten . Confinement in this model was also shown to be
accompanied by monopole condensation \gross , as expected.

Moreover, by adding $N_f$ D6 brane probes to this D4
brane supergravity
background, Kruczenski, Mateos, Myers and Winters \KruczenskiUQ\
recently constructed a string model dual to four dimensional QCD with $N$
colours and $N_f$ flavours of mass $m_q$.
They have shown that for $m_q=0 $ the model exhibits
spontaneous chiral symmetry
breaking by a quark condensate, extending previous results
on chiral  symmetry
breaking condensates found in a different  gravity dual model in \chir .
{}For $m_q>0$ the pion mass exactly satisfies
the Gell-Mann-Oakes-Renner formula which relates it to the quark
condensate.
In addition, the model provides a simple geometric interpretation of the
Vafa-Witten theorem \KruczenskiUQ ,
 and the authors of \barbon\  found evidence 
that  the Veneziano-Witten theorem
(relating
the mass of the $\eta ' $ to the topological susceptibility computed
in \hashi ) is satisfied  (the Veneziano-Witten formula
has  been derived in a setup
 of D3 branes on orbifold singularities in \armoni ). 
The model was also used to describe
the recently observed pentaquarks \bando ,
obtaining reasonable predictions for masses.

Computing the full string spectrum in this string model of large 
$N$ QCD may give
important insights on  general features of the 
glueball spectrum of large $N$ QCD
and on what are the scales associated with the Kaluza-Klein
particles  of the model. The determination of the string spectrum in a
curved background is in general a  complicated problem. 
However, as suggested in \GKP ,
one can explore the spectrum at large quantum numbers by considering
classical string configurations as well as quantum fluctuations about it.
This approach led to remarkable results in the $AdS_5\times S^5 $ case
(see e.g. \Spins\ and references therein).
In this paper we will consider a large variety of 
classical string configurations with the aim of determining the
typical energy scales at strong $\lambda $ coupling 
of physical string states with different quantum numbers.\foot{
Previous discussions on semiclassical string solutions in this model
are in \refs{\barb,\AlishahihaVI }. 
Studies of semiclassical string solutions in other confining
backgrounds can be found in \MN .}
We will find a universal pattern, indicating that for any string solution
the derivative of the ratio 
${\rm Energy}/ \sqrt{T_{\rm YM}}$ with respect to $\lambda$
($T_{\rm YM}$ is the string tension) is negative or vanishing,
but never positive. In other words, to leading order at $\lambda \gg
1$, the energies of the states in units of the string tension 
increase or remain 
constant as $\lambda $ is decreased.
Some
states become unstable when the angular momentum is lower than some
critical value. Remarkably, this occurs precisely at the point where
the derivative of  ${E\over \sqrt{T_{\rm YM}}}$ with respect to
$\lambda $ becomes positive, thus confirming the rule 
that, in units of the string tension, 
the energies of the  states 
increase  or remain 
constant in decreasing $\lambda $ (in the region $\lambda \gg 1$).
We also find some string solutions which exist only for special values
of the coupling $\lambda $.


\newsec{QCD$_{3+1}$ model: generalities and restrictions}

The background of the Witten QCD model can be obtained
from the euclidean non-extremal D4 brane
(we follow the notation of \russo ),
\eqn\metric{\eqalign{ds_{10}^2 = \alpha '
\bigg( {8\pi \lambda\over 3}{u^3\over u_0}
\big(dx_1^2 +...+dx_4^2\big)+ {8 \pi \lambda\over 27}
\big({u\over u_0}\big)^3
h(u) d\theta_1^2 +{8 \pi \lambda\over 3} 
{du^2\over u\, u_0
h(u)} +{2 \pi \lambda\over 3}{u\over u_0} d\Omega_4^2\bigg) ,
}}
$$
e^{2\phi } = {8\pi \over 27} {\lambda ^3u ^3\over u_0^3}\ {1\over
N^2}\ , \ \ \  \ \ h(u) \equiv  1 - {u_0^6\over u^6}\ ,
$$
\eqn\ady{d\Omega_4^2= d\alpha^2 +
\cos^2\alpha\left( d\theta^2 + \cos^2\theta\,d\phi^2 +
\sin^2\theta\,d\varphi^2 \right)\,,
}
where 
$\lambda $ is the `t
Hooft coupling,
$2 \pi \lambda =  g^2_{\rm YM} N$.
The Euclidean time is $t_E=R_0\theta_1 $, $R_0=(3u_0)^{-1}$,
and   $\theta_1$ is $2\pi $ periodic.
 To describe the Minkowskian QCD$_{3+1}$ theory, one  makes the Wick
rotation $x_4\to i x_0$.
The metric exhibits a coordinate singularity at $u=u_0$, which can be
removed as usual  by going  to e.g. Rindler-type coordinates
$$
{u^3\over u_0^3}=1+{\rho^2\over \rho_0^2}\equiv h_1(\rho )\ ,\ \ \ \
\rho_0^2\equiv u_0^3\ ,\ \ \ \ \ \ \rho \in [0,\infty)\ .
$$
One finds
$$
ds_{10}^2  = \alpha' \bigg({8\pi \lambda  u_0^2 \over 3}
h_1
\big(-dx_0^2+dx_1^2 +...+dx_3^2\big)+ {16 \pi \lambda\over 27}
{h_2\over h_1} \, {\rho^2\over \rho_0^2} \,d\theta_1^2
$$
\eqn\rind{
+ {16 \pi \lambda\over 27\rho_0^2} \,
{h_1^{1/3} \over h_2} d\rho^2  +{2 \pi \lambda\over 3}h_1^{1/3}
d\Omega_4^2\bigg)\,.
}
$$
e^{2\phi } ={8\pi \over 27} {\lambda ^3 h_1}\ {1\over
N^2} \ ,\ \ \ \ \ h_2\equiv 1+{\rho^2\over 2\rho_0^2}\ .
$$
In this model, QCD arises as the dimensional reduction of five
dimensional Yang-Mills theory, which is compactified on a circle of
radius $(3u_0)^{-1}$ with anti-periodic boundary conditions for the fermions.
From the point of view of the gauge theory,
the parameter $u_0$ plays the role of an ultraviolet cutoff,
and the coupling $g_{\rm YM}$ should be understood as the bare
coupling at distances corresponding to $1/u_0$.
There are Kaluza-Klein particles with masses of order $O(u_0)$.
To obtain $d=3+1$ QCD, one should take $u_0\to\infty $ and $\lambda
\to 0$ with a fixed string tension \refs{\witten,\gross}, 
which for small $\lambda $ should
be of the form
$T_{\rm YM}\sim \exp(-2 b/\lambda )u_0^2\sim \Lambda_{\rm QCD}^2$.

In QCD, the  glueball mass scale is $\Lambda_{\rm QCD}$, so it is  
essentially set by the square root of the string tension. 
In the present supergravity background,
there are  glueball states whose masses
are proportional to $u_0$ for  $\lambda\gg 1$, with a coefficient
independent of $\lambda $, which
originate from the supergravity sector of the string spectrum,
and, as we shall see, glueball states of masses proportional to 
the square root of the string tension,
which at $\lambda\gg 1$ is given by
\eqn\tty{
T_{\rm YM}={4\over 3} \lambda u_0^2\ ,\ \ \ \ \lambda\gg 1\ .
}
This is the case for highly excited string states, as in the example
shown in the next section. 
This tension derived from the energy formula (3.8) agrees with the tension
obtained from the Wilson loop
calculation (see  \refs{\witten,\BrandhuberER,\gross}).
Thus the masses of these two sets of states differ by a
factor $\sqrt{\lambda }$.
Since in pure gluodynamics there is only a single mass scale, one expects
that the  states  surviving
the limit  $\lambda\ll 1$ will all have masses of the
same order, proportional to the square root of the string tension.

To take the small $\lambda $ limit in the geometry, one defines
$u'=\lambda u/u_0$ so that the dilaton coupling \metric\ is
fixed in taking the limit $\lambda\to 0$ at fixed
$u'$. This gives the extremal D4 brane, which is singular at $u'=0$.
In this region, the supergravity approximation cannot be justified.
The gravity solution \metric\ solves the leading $\alpha' $ order string
equations, and therefore it applies as long as further $\alpha' $
corrections are suppressed.
To ensure that these corrections
are small one is to demand that  curvature invariants
and higher derivatives of the dilaton are small in string
units. In the present case,
these are all suppressed by inverse powers of $\lambda $. In particular,
the curvature scalar is
$\alpha ' {\cal R} = -{27\over 8 \pi \lambda}
{u_0\over u} (5 -{u_0^6\over u^6})$, which is small in all the space
provided $\lambda \gg 1$.

In addition to the requirement of large $\lambda $,
one must require that the string coupling constant is small,
so that string loop corrections are suppressed. 
This implies the condition $u<u_{max}$, with
$u_{max} \approx {N^{{2\over 3}} \over \lambda} u_0$.
By taking $N$ sufficiently large, one can extend the applicability
of the tree-level approximation to arbitrary large values of $u$.

\newsec{ Semiclassical string spectrum}

In the physical string spectrum, there are basically two types of
string states:

\smallskip
\noindent 1) The states having zero quantum numbers in $S^1$ and $S^4$, which
in the small $\lambda $ limit should correspond to glueballs of QCD.
\smallskip

\noindent 2) The ``Kaluza-Klein'' type of string states,
having non-zero quantum numbers in $S^1$ or in $S^4$.
Such states should decouple in the small $\lambda\ll 1 $ limit since
they have no counterpart in ordinary QCD.

\smallskip

In this section we  compare the
masses of semiclassical  string states describing glueball states
and Kaluza-Klein states.
We will argue that they correspond to two
different mass scales which get
separated as $\lambda $ is decreased,
giving rise to heavy Kaluza-Klein states and lighter glueball states.

The states are characterized by angular momentum quantum numbers,
which are conserved, and by  winding numbers of the strings 
around $S^1$ and $S^4$
--or number of foldings of a folded string-- which are
not conserved in interactions, because all circles are contractible. 
Nevertheless, they can be used
to characterize the string states of the free string theory.


\medskip

In order to obtain the relations between the energy and the angular
momenta
of classical string configurations,
one can start with the Polyakov action in the conformal gauge,
\eqn\nambugoto{ I = -{1\over 4\,\pi\alpha '}\int\,d\tau\,d\sigma\,
G_{\mu\nu}\,\partial_\alpha
X^\mu(\tau,\sigma)\, \partial_\beta X^\nu (\tau,\sigma)\,\eta^{\alpha\beta}\,.
}
The classical strings must satisfy the equations of motion
\eqn\eom{ \left( -{\partial G_{\rho \nu}\over
\partial X^\mu}
+2 {\partial G_{\mu\nu}\over\partial X^\rho} \right)\,
\left(\dot X^\rho\, \dot X^\nu -  {X^\rho}'\, { X^\nu}'\right)
+2\,G_{\mu\nu}\,\left( \ddot X^\nu - {X^\nu}'' \right)=0\,,
}
where dots and primes denote derivatives with respect to $\tau $ and
$\sigma $, respectively.
In addition, one has the usual Virasoro constraints
\eqn\V{ G_{\mu\nu}({\dot X}^\mu{\dot X}^\nu+X'^\mu X'^\nu )=0\,,
}
\eqn\U{
 G_{\mu\nu}{\dot X}^\mu X'^\nu=0\,.
}
{}For a diagonal target metric, as in the present case,
the energy and the angular momentum 
in a generic angle $\varphi $ are given by
\eqn\enes{
E=-{1\ov 2\pi\a '}\int _0^{2\pi} d\s\ G_{00 }(X) \partial_\tau X^0\ ,
\ \ \ \
J_\varphi =
{1\ov 2\pi\a '}\int _0^{2\pi} d\s\ G_{\varphi \varphi }(X) \partial_\tau
\varphi \ .
}
Combining these equations, below we find  $E=E(J_{\varphi_i} )$ for different
configurations. 

\subsec{Glueball states and the Regge string}

In comparing masses of glueballs to masses of
Kaluza-Klein states, in this section 
we will only consider classical string states
of high spin. 
Supergravity glueballs will be discussed in section 5. 
We stress that, in general, the calculation of masses at
$\lambda\gg 1$ based on the geometry \metric\
 cannot be extrapolated to the weak coupling regime
$\lambda\ll 1$ (see section 2 and also discussion in section 4).

\medskip

To look for classical string solutions, it is convenient to use
the radial coordinate $\rho $ which is regular at the horizon located
at $\rho=0$
(in working in the singular $u$ coordinate frame one misses some solutions of
strings lying on the horizon).

To describe  a string rotating in $\Re^3$, we use cylindrical coordinates
$dx_1^2+dx_2^2+dx_3^2=dr^2+r^2 d\beta^2+dx_3^2$, and consider
the following   configuration:
\eqn\along{t=\kappa\,\tau\,,\quad\rho=0\,,\quad
r=r(\sigma)\,, \quad\beta=\omega\tau\,.
}
It represents a folded closed string located at the horizon which 
rotates in $\Re^3$.
The action becomes
\eqn\actionregge{I = - {1\over 4 \pi } {8 \pi \lambda \over 3}  u_0^2
\int d\tau d\sigma (\kappa^2 - r^2 \omega^2 + r'^2)\,.
}
The  equations of motion are consistent with those of the original
action.
The equation of motion for the variable $r$
determines $r= r_0 \sin(\omega \sigma)$. The periodic boundary
condition
for $r$ then implies that $\omega $ must be an integer.
We shall consider $\omega=1$, which corresponds to the minimal 
one-winding case with
the two foldings of the string located at $r=r_0$
(the string
rotates around its center of mass at $r=0$).
The Virasoro constraint \V\ imposes $\kappa^2 = r_0^2$.
Then, using \enes , we find
that  the energy $E$ and the
angular momentum $J_{\beta}$ are related by
the Regge relation
\eqn\regge{
E_{\rm regge} = 4 u_0 \sqrt{{\pi  \over 3 }\, \lambda \, J_{\beta}} \, ,
}
corresponding to the string tension \tty . This relation was pointed
out
in this context in \barb . 
{}For a string with a generic integer $\omega\neq 1$, the energy
formula \regge\ is multiplied by a factor $\sqrt{\omega }$.

In order to compare the different energy formulas, in what follows
it will be useful to consider the ratio with the string tension, which
is the natural scale in pure QCD. From \tty\ and \regge\ we obtain
\eqn\regg{
{E_{\rm regge}\over \sqrt{T_{\rm YM} } } =2\sqrt{\pi J_\beta }\ .
}

The above example of the Regge string illustrates an important
feature of the string spectrum. Namely, all glueball states
corresponding to strings lying at $\rho=0$ have the same behavior
$E_{\rm glueball}/ \sqrt{T_{\rm YM} }  ={\rm const.} $ The
reason is that once we set $\rho=0$ the string configurations on
$\Re^3$ are the same as in flat space upon changing the flat space
string tension $T$ by $T_{\rm YM}$ \tty . Since this is the only
dimensionful scale and the parameter $\lambda $ does not appear
independently, the energies for all these strings must be
proportional to $\sqrt{T_{\rm YM} }$.
This means that in units of $T_{\rm YM}$, to leading order in
the expansion in $1/\lambda $, the energies
of strings lying at  $\rho=0$  do not change as $\lambda $ is decreased.
This is in contrast to some strings with Kaluza-Klein quantum numbers
examined below, and
 in contrast with supergravity glueballs (see section 5),
which have a $1/\sqrt{\lambda }$ leading behavior.

The strings lying at $\rho=0$ cover a significant part of the glueball
string spectrum; besides the special Regge string pointed out above
one can  write the full flat space spectrum of four dimensional
strings. The glueballs that are not included are those where
the string also fluctuate in the radial direction $\rho $. This case
is examined in section 5, and the energy exhibits a different
dependence on $\lambda $.

\subsec{String spinning in the cigar}

We start with the Nambu-Goto action and consider the ansatz.
\eqn\jhh{
t=\tau\ ,\ \ \ \ u=u(\sigma )\ ,\ \ \ \ \theta_1=\omega_1\tau \ .
}
This represents a folded string spinning in the ``cigar'' geometry described
by $\rho , \ \theta_1 $ (or $u,\ \theta_1$).
This configuration first appeared in \barb . Here we are interested in
the dependence of the energy as a function of $\lambda $.
The configuration has a non-trivial quantum number, the angular momentum
$J_{\theta_1}$ in the internal
coordinate $\theta_1 $, and therefore is of Kaluza-Klein type.

The string has energy and angular momentum  given by
\eqn\ftf{
E_{\rm KK}={8\lambda \over 3} u_0 n \sqrt{ y_m^{3}-1}\int _1^{y_m} dy
{y^3\over \sqrt{(y_m^3-y^3)(y^3-1)} }\ ,\ \ \ \ \ 
y_m^3\equiv {\omega_1^2\over \omega_1^2-9u_0^2}\ ,
}
\eqn\wui{
J_{\theta_1}= {8\lambda \over 9} n y_m^{3/2}\int _1^{y_m} dy
{\sqrt{y^3-1}\over \sqrt{y_m^3-y^3} }\ .
} 
Here $n$ is an integer denoting the number of foldings (windings)
of the string. The string is extended from $u=u_0$ to $u=u_{\rm
max}=u_0\sqrt{y_m}$ (the center of mass being at $u=u_0$). 
If $\omega_1\to 3u_0$, then 
$u_{\rm max}\to\infty $.
These equations define $E_{\rm KK}=E_{\rm KK}(J_{\theta_1},n, \lambda )$
in a parametric way. The numeric plot shows that
${E_{\rm KK}\over \sqrt{T_{\rm YM}}}$ is a monotonically decreasing
function of $\lambda $. This seems to be a universal behavior that will
hold for all stable configurations examined below.
 

\subsec{Strings on $\Re^3\times S^4$}

\noindent {\it i)} The simplest state with $S^4$ charge is given 
by a point-like string
rotating at the equator of $S^4$. It is described by
\eqn\particle{t= \kappa\, \tau\,,\quad\alpha=0\,,\quad\theta={\pi\over 2}\quad
{\rm and}\quad \varphi=\omega\, t\,.}
Now we find that the energy is given by
\eqn\eparticle{
E_{\rm KK}= 2\, u_0\,J_\varphi\,,
}
so that
\eqn\epar{
{E_{\rm KK}\over \sqrt{T_{\rm YM} } } =
{1\over \sqrt{\lambda }}\ \sqrt{3}J_\varphi \ .
}
Comparing the $\lambda $ dependence of the energies \eparticle\ and
\regge , corresponding to the
Regge string rotating $\Re^3$, we see that
the latter has an extra factor of $\sqrt{\lambda}$.
As mentioned above, as $\lambda $ is decreased (while keeping a large value $\lambda\gg 1$ so
that the supergravity approximation still applies) we see that
${E_{\rm regge}\over \sqrt{T_{\rm YM} } }$ remains constant and
${E_{\rm KK}\over \sqrt{T_{\rm YM} }}$ increases (in this case  
like ${1\over \sqrt{\lambda }}$).

\medskip

\noindent {\it ii)} One can also write down a solution describing 
strings spinning on $S^4$. Now we use the Nambu-Goto action and consider
the  ansatz
$$
t=\tau \ ,\ \ \ \ \rho=0\ ,\ \ \ \ \alpha=0\ ,
\ \ \ \theta = \theta (\sigma )\ ,\ \ \ \ \varphi =\omega
\tau \ .
$$
This is essentially the same solution appearing in $AdS_5\times S^5$
\GKP , given in terms of elliptic functions.
The equation for $\theta $ is that of a pendulum in a
constant gravitational field. There are folded solutions (representing
a string extended in the $\theta $ direction from $\theta =0$ to some
maximum value $<\pi/2 $)
 and circular
solutions
(strings winding around the $\theta $ direction of $S^4$).
For  folded strings, there are two limits, short and long strings.
In the case of long strings,
one obtains, similarly as in
\GKP , $E_{\rm KK}\cong 2 u_0 J_\varphi $.
For short strings, one obtains
$E_{\rm KK}\gg  u_0 J_\varphi $. 
For fixed $J_\varphi $, varying $\lambda $ from 0 to $\infty $
interpolates
between the long and short string regime.
One can show numerically that 
the derivative of the ratio ${E_{\rm KK}\over \sqrt{T_{\rm YM} }}$ 
with respect to $\lambda $ is negative for all $\lambda $.

In the case of strings winding in the $\theta $ direction we find
\eqn\enpend{
E_{\rm KK}= {4 \lambda \over 3} n \,u_0 \int_0^{2\pi} d\theta\,
{1\over \sqrt{1-\eta^2 \sin^2\theta}}\,, \quad J_\varphi= {2 \lambda \over 3} n \,\eta 
\int_0^{2\pi} d\theta\,
{\sin^2\theta\over \sqrt{1-\eta^2 \sin^2\theta}}\,,
}
where $\eta={\omega\over 2 u_0}< 1$ and $n$ is the winding number,
$\theta(\sigma +2\pi )=\theta (\s )+2\pi n$. These equations 
define $E_{\rm KK}(J_\varphi,n, \lambda )$. 
The equations can also be written as
$E=E(J_\varphi,n,\eta )\ ,\ \lambda =\lambda(J_\varphi,n,\eta )$.
Fixing $J_\varphi$ and $n$ one can plot the energy as a function of
$\lambda $.
A plot of  ${E_{\rm KK}\over \sqrt{T_{\rm YM} }}$ 
as a function of $\lambda $ shows that for $\lambda $ higher than some
 critical value
the derivative of ${E_{\rm KK}\over \sqrt{T_{\rm YM} }}$ becomes positive.
The transition occurs at a point $\eta_1\cong 0.909$, with $\eta_1$
being the root of the equation $K(\eta^2)=2E(\eta^2)$, $K$ and $E$
being the standard elliptic functions.
The critical $\lambda $ is then 
$\lambda_1=\lambda(J_\varphi,n,\eta_1 )$. 
The derivative of ${E_{\rm KK}\over \sqrt{T_{\rm YM} }}$ is positive
for $\eta<\eta_1 $, which corresponds to $\lambda>\lambda_1$.

On the other hand, the string becomes unstable 
for an angular momentum below
a certain critical value (for fixed $J_\varphi $, 
the string is unstable for  $\lambda $ higher than some critical
value).
This instability is easy to understand 
in the extreme limit $\omega=0$; one has a winding string
with no angular momentum, and a small perturbation
makes the string to slide off the equatorial plane of the sphere
due to the string tension.
To determine the critical value, one must
look at the Lagrangian of small fluctuations. 
Expanding 
$\alpha = 0 +\tilde\alpha (\theta, t)$ 
we find
$$
L_{\rm fluc}\cong{1\over u_0}{1\over \sqrt{1-\eta^2\sin^2\theta }}
\ \dot {\tilde\alpha}^2+ 4 u_0 
{1-2\eta^2\sin^2\theta \over \sqrt{1-\eta^2\sin^2\theta }}\ 
{\tilde\alpha}^2-  4 u_0 
 \sqrt{1-\eta^2\sin^2\theta }\ {\tilde\alpha} ^{'2} \,.
$$
In the case $\omega=0$, i.e. $\eta=0$, the instability is evident
since the potential is negative, and one gets harmonic modes with
imaginary frequencies.
For $\eta \neq 0$, an exact study of the stability requires solving
a coupled system of an infinite set of harmonic oscillators.
Nevertheless, one can probe the system by a small perturbation at
$\tau=0$
of the
form ${\tilde\alpha} (\theta ,0)=\delta_0 $.
Then one sees immediately that the potential becomes negative
precisely at $\eta<\eta_1\cong 0.909$ , and the equation for the zero mode at
small times is that
of a harmonic oscillator with imaginary frequency, indicating
instability. 
Thus instabilities seem to appear precisely at
the same point where the derivative of ${E_{\rm KK}\over \sqrt{T_{\rm
YM} }}$ with respect to $\lambda $
becomes  positive.
This remarkable feature is seen more clearly in other solutions
examined below. Namely, the solutions become unstable precisely 
in the same interval
where the derivative of the ratio ${E_{\rm KK}\over \sqrt{T_{\rm YM} }}$ 
with respect to $\lambda $ is positive. The instability is of the same
type, there is a winding string that slides off a big circle of the
sphere
when the angular momentum is lower than some critical value.

\medskip

\noindent
{\it iii)} The string spins on the $\Re^3\times S^4$ space
\eqn\squrtr{
\rho=0\ ,\ \ \ t=\kappa\,\tau\,,\ \ \ r=r_0\ ,\ \ \ \beta=m\,(\sigma+\tau)\,,
\quad\alpha=0\,,\quad\theta={\pi\over 2}\,,\quad
\varphi=-n\,\sigma+\omega\tau\, .
}
The Virasoro constraint \U\ implies
the relation
$ n J_\varphi = m J_\beta
 $
between the angular momenta associated with the angles $\varphi$ and $\beta$.
Now we find
$$
4u_0^2r_0^2 m^2=n\omega \ ,\ \ \ \ 
J_\varphi= {2\pi \lambda\over 3}\ \omega \ ,
$$
\eqn\esqurtr{
E_{\rm KK}=2\,u_0\,{J_\varphi }+{4\,\pi \lambda\over 3} u_0\,n\, ,
}
i.e.
\eqn\esqur{
{E_{\rm KK}\over \sqrt{T_{\rm YM} } } ={\sqrt{3}\over \sqrt{\lambda }}
\ \,{J_\varphi }+{2\,\pi \sqrt{\lambda }\over \sqrt{3}} \,n\,.
}
As $\lambda $ is decreased, the ratio $E_{\rm KK}/ \sqrt{T_{\rm YM}} $
\esqur\ of these
states first decreases and then increases.
The transition is at $\lambda_0= {3 J_\varphi\over 2\pi n}$.
For states with $\lambda_0={3 J_\varphi\over 2\pi n}\gg 1$, 
the energy formula applies
in the region $\lambda<\lambda_0$, so they exhibit
the behavior ${E_{\rm KK}\over \sqrt{T_{\rm YM}}} \sim {1\over
\sqrt{\lambda }}$.
On the other hand, for states with ${3 J_\varphi\over 2\pi n}\ll 1$,
the formula \esqur\ cannot be extrapolated into the region
$\lambda<\lambda_0$.


Remarkably, these string  configurations become classically
unstable precisely when
$ J_\varphi < {2\pi \lambda\over 3}\ n$, i.e. precisely when the derivative of
${E_{\rm KK}\over \sqrt{T_{\rm YM} } } $ with respect to $\lambda $
is positive.
This can be seen by examining quadratic fluctuations in the $\theta $
direction, $\theta={\pi\over 2}+\tilde\theta (\tau )$.
We get the Lagrangian
$$
L_{\rm fluc} \cong \dot {\tilde \theta}^2 - (\omega^2 -n^2) {\tilde
\theta}^2\ ,
$$
so there is an instability mode for $\omega < n$. Since 
$J_\varphi= {2\pi \lambda\over 3}\ \omega $, 
this corresponds to $ J_\varphi < {2\pi \lambda\over 3}\ n $, i.e.
the string is unstable when
$\lambda >\lambda_0$. The instability mode can be understood
as a tendency of the circular string wrapping the $S^4$ to slip off the side
when the angular momentum is insufficient to keep it stable.

Thus, for all stable string configurations of this type
$E_{\rm KK}/ \sqrt{T_{\rm YM}} $  increases in decreasing $\lambda $.
In particular,
{} for states with $J_\varphi \gg n$, the increasing regime already
starts at $\lambda \gg 1$, so their energy reproduces the ${1\over
\sqrt{\lambda }}$ behavior of  previous cases.


\subsec{String on $\Re^3\times S^1$}

The configuration
\eqn\srsun{t=\kappa\,\tau\,,\quad u=u_1\,, \quad\theta_1=n_1\,
\sigma-\omega_1\,\tau\,,
\quad r= r_0\,,\quad
\beta=m\,(\sigma+\tau)\,,
}
with $\omega ,n,m >0$,
solves the equations of motion and Virasoro constraints with
$({u_1\over u_0})^6 = {\omega_1 \over n_1} $
and with the relation between the parameters
$\omega_1-n_1= (3 m u_0 r_0 )^2/n_1\ $ ($n_1$ and $m$ are integers).
The energy of the configuration is
\eqn\feng{
E_{\rm KK}= {8 \pi \lambda  \over 9} u_0\,n_1({\omega_1\over n_1} +
1)\,\sqrt{{\omega_1 \over n_1}-1}\ ,
}
whereas the (integer) momentum  $p_{\theta_1}$
over $S^1$ is given by
\eqn\momm{
p_{\theta_1} = {8 \pi \lambda  \over 27}n_1({\omega_1\over n_1} - 1)
\sqrt{{\omega_1 \over n_1}}\ .
}
{}From the constraint \U\ one has the relation
${J_{\beta} m}= {p_{\theta_1} n_1}\, $.

The $\lambda $ dependence of the energy in \feng\ is hidden in
$\omega_1=\omega_1(\lambda)$, defined in terms of the quantized momentum
$p_{\theta_1}=0,\pm
1,\pm 2,...$ and
$\lambda $ through \momm . One can write explicitly
 $\omega_1=\omega_1(\lambda)$ by
solving the cubic equation \momm . 
The function $\omega_1(\lambda)/n_1$ tends to infinity at $\lambda=0$
and it approaches $1$ at $\lambda=\infty$.
The energy  ${E_{\rm KK}(\lambda)\over \sqrt{T_{\rm YM} }}  $ goes
like $\sim {1\over \sqrt{\lambda }}$ near $\lambda=0$ and 
approaches a constant at $\lambda=\infty$.
This can be seen
explicitly by considering limits.
In the case of large $\lambda$, 
${\omega_1\over n_1} - 1 \ll 1$ we find\foot{Note that, since
$n_1 \gg \omega_1 - n_1 \cong {27\over 8 \pi } {p_{\theta_1}\over \lambda}$,
it follows that  $\lambda \gg p_{\theta_1}/n_1$ and
therefore $E_{\rm KK} \gg u_0 p_{\theta_1}\ .$
Since $ p_{\theta_1} $ is quantized, this implies $E_{\rm KK} \gg u_0$. 
}
\eqn\qwer{
E_{\rm KK}\cong 4 \sqrt{2\pi\lambda\over 3} \sqrt{ p_{\theta_1} n_1}\ u_0\ ,
}
{}For  $\lambda \ll p_{\theta_1}/n_1$
one has  ${\omega_1\over n_1} \gg 1$
and
we obtain $E \cong 3 u_0  p_{\theta_1}$, showing that 
${E_{\rm KK}\over \sqrt{T_{\rm YM} }} \sim {1\over \sqrt{\lambda }}$
for $1\ll {\lambda}\ll  p_{\theta_1}/n_1 $.

\subsec{String spinning on $S^1\times S^4$}

\noindent
{\it i)}  We now consider
the configuration:
\eqn\sunsqur{t=\kappa\,\tau\,,\quad u=u_1\ ,\quad
\theta_1=\omega_1\,\tau-n_1\,\sigma\,,
\quad\alpha=0\,,\quad\theta={\pi\over 2}\,,\quad\varphi=m \, (\sigma+\tau)\,,
}
with $w_1,n_1,m>0$.
For this
configuration
$$\big({u_1\over u_0}\big)^6 = {\omega_1\over n_1}\ \ ,\ \ \ \
\kappa^2 ={(\omega_1-n_1)(\omega_1+n_1)^2 \over 9 u_0^2 \,\omega_1}\ .
$$
Note that the configuration only exists for
$\omega_1 > n_1$, i.e. for $u_1>u_0$.
After some straightforward
manipulations, one gets the following formulas for the energy and the
 momenta:
\eqn\otraq{
E_{\rm KK}={8\pi\lambda \over 9} u_0 
\sqrt{ {\omega_1\over n_1}-1} \big( \omega_1
+n_1 \big) \ ,\ \
}
\eqn\jkl{
 p_{\theta_1} =  {8 \pi  \over 27}\lambda\,(\omega_1-n_1)
\sqrt{\omega_1\over n_1}\,,
\ \ \ J_\varphi = {4 \pi  \over 9} \lambda\,({\omega_1\over n_1})^{1 \over
3}\sqrt{n_1(\omega_1-n_1)}\ ,
}
\eqn\uio{
m={2\over 3} ({\omega_1\over n_1})^{1 \over
6}\sqrt{n_1(\omega_1-n_1)}\ .
}
Note that ${J_{\varphi} m}= {p_{\theta_1} n_1}$.
These relations are similar to the ones appearing for the
configuration of sect. 3.4, except that the solution of section
3.4 has the additional parameter $r_0$.

This string solution has the remarkable 
property that it exists only for a special value of $\lambda $.
Indeed, fixing four integer quantum numbers satisfying
 ${J_{\varphi} m}= {p_{\theta_1} n_1}$, 
then the two equations~\jkl\ determine $\omega_1$ and $\lambda $ in
terms
of integer numbers. The equation \uio\ is then satisfied automatically.
For a small variation
of $\lambda $,  this string state ceases to be physical and  disappears
from the string spectrum.
In other words, there is a set of special values of $\lambda $
(parametrized by integers $J_\varphi , \ p_{\theta_1}, \ n_1$)
for which the string spectrum has extra states.

Explicit formulas for the energy of the state in terms of the
quantum numbers can be obtained in different limits.
In the limit ${\omega_1\over n_1}-1 \ll 1$,  the energy becomes
\eqn\esunsqutra{E_{\rm KK}\cong 4\,u_0\,J_\varphi\,.
}
In the opposite limit ${\omega_1\over n_1} \gg 1$, we obtain
\eqn\esunsqutrb{
E_{\rm KK}\cong 3\,u_0\,p_{\theta_1}\,.
}

\bigskip
\noindent
{\it ii)}
Here
we use the alternative parametrization of $S^4$,
\eqn\ali{d\Omega^2_4= d\gamma^2 + \cos^2\gamma\, d\varphi_1^2 +
\sin^2\gamma \left( d\psi^2+\cos^2\psi\,
d\varphi_2^2\right)\,.
}
Now the configuration considered is
\eqn\sali{
t= \kappa \tau\,,\quad u =u_1\,,\quad \theta_1=\omega_1 \tau \,,
\quad \gamma=m \sigma\,,
\quad\psi= 0 \,,\quad \varphi_2 = \omega \tau\,,
\quad \varphi_1 =\omega \tau\, .
}
{} This configuration appeared in \AlishahihaVI .
From the equations of motion and the Virasoro constraints we get
\eqn\xsol{
({u_1\over u_0})^4 =  {4 \omega_1^2\over  6m^2+ 3\omega^2}\ ,\ \ \ \
\kappa^2 ={8\omega_1^3 +3\sqrt{3}(m^2 + 2 \omega^2)\sqrt{2m^2
+\omega^2}\over 72 u_0^2 \omega_1}\,.
}
The particular case considered in \AlishahihaVI , section 6, corresponds to
$u_1=u_0$, which implies
$4 \omega_1^2 = 6m^2+ 3\omega^2$ and then $\kappa^2= {{m^2 + \omega^2}\over
4u_0^2}$. The horizon limit 
$u_1 \rightarrow u_0 \,(\rho \rightarrow 0)$ is smooth for
this configuration.

The energy, the angular momentum 
$J_{\varphi}:=J_{\varphi_1}= J_{\varphi_2}$ and the
momentum $p_{\theta_1}$ are, respectively
\eqn\buo{
E_{\rm KK}= {8\pi \over 3} \lambda \ {u_1^3\over u_0}\ \kappa\ ,\ \ \ \
J_{\varphi}= {\pi  \over 3}\lambda\, \omega \,{u_1 \over u_0}\,,\ \ 
\ \quad
p_{\theta_1}= {8\pi  \over 27}\lambda\,
({u_1 \over u_0})^3 \,\big(1- ({u_0 \over u_1})^6\big)
\,\omega_1\,.
}
For a string lying near the horizon,  ${u_1 \sim u_0}$, the
energy becomes
\eqn\gre{
E_{\rm KK}\cong 4u_0\sqrt{J_\varphi^2+ \big({\pi\over 3}\lambda
m\big)^2} \ .
}
Hence
\eqn\kkali{
{E_{\rm KK}\over \sqrt{T_{\rm YM} }}\cong
2\sqrt{3}\sqrt{{J_\varphi^2\over \lambda}+
\big({\pi\over 3} m\big)^2\lambda } \ .
}
Similarly to the discussion in section 3.2, we note that,
as $\lambda $ is decreased, the ratio $E_{\rm KK}/ \sqrt{T_{\rm YM}}$
of these states first decreases and then increases.
The transition is now at $\lambda_0= {3 \over \pi m} J_\varphi$.
For states with $\lambda_0={3 \over \pi m} J_\varphi\gg 1$, the energy formula
applies
in the region $\lambda<\lambda_0$, so they exhibit
the behavior ${E_{\rm KK}\over \sqrt{T_{\rm YM}}} \sim {1\over
\sqrt{\lambda }}$ of the previous KK configurations.
On the other hand, for states with ${3 \over \pi m} J_\varphi\ll 1$,
\kkali\ cannot be extrapolated into the region
$\lambda<\lambda_0$.

Just as the strings ii) and iii) of section 3.3, these string
configurations become classically
unstable precisely when
$ J_\varphi < {\pi \over 3}\lambda m$, i.e. when the derivative
of  ${E_{\rm KK}\over \sqrt{T_{\rm YM}}}$ with respect to $\lambda $
is positive.
To  identify instability modes it is
convenient to examine the configuration in terms of
the coordinates \ady ,
\eqn\alibis{
\alpha=0\,,\quad\theta={\pi\over 4}\,,\quad\phi=\omega \tau+m\sigma\,,
\quad\varphi= \omega \tau-m\sigma\,.
}
Then the instability shows up by examining quadratic fluctuations in the $\alpha$
direction. We get the Lagrangian
$$
L_{\rm fluc}\cong \dot {\tilde \alpha}^2 - (\omega^2 -n^2) {\tilde
\alpha}^2\ ,
$$
exhibiting  an instability mode precisely for $\omega < n$, i.e. for
 $ J_\varphi < {\pi \over 3}\lambda m$.

Now consider  $u_1/u_0 \gg 1$. In this case
the energy comes mostly from the motion in $\theta_1 $  and from \xsol
, \buo\ we get
\eqn\masen{
E_{\rm KK} \cong 3 u_0 p_{\theta_1 }\,.
}
Thus for $u_1/u_0 \gg 1$ one has the behavior
${E_{\rm KK}\over \sqrt{T_{\rm YM} }} \sim {1\over \sqrt{\lambda }}$.


\bigskip

There are also  circular string configurations similar to those studied
in \ART . In the present case the circular string lies on the horizon
at $\rho=0$,
and the solutions are essentially those of \ART .
The energy has a similar structure as \gre\ and in the regime
$\lambda\ll J/m$, where the angular momentum term dominates, 
it  has the behavior
${E_{\rm KK}\over \sqrt{T_{\rm YM} }} \sim {1\over \sqrt{\lambda }}$.

\newsec{ A small slice of the $\lambda \ll 1 $ spectrum}

As pointed out in section 2, the scalar curvature  of the geometry
\metric\  is small in units of $\alpha '$ provided
\eqn\qqp{
{u_0\over \lambda u}\ll 1\ .
}
One can check that this condition ensures that all curvature
invariants are small. It also ensures that higher derivatives of the dilaton
field are small.
By considering strings moving on the region $u\gg u_0/\lambda $, one
can therefore
describe the geometry in terms of  the leading order metric \metric ,
since in this region higher order $\alpha '$
corrections can be neglected.
For such strings, corrections to the  classical energy
formula could
therefore only originate from quantum sigma model corrections, but not from
corrections to the background.
However, assuming as usual that the classical limit exists, quantum
corrections to the energy can also be neglected relative to the
leading term  by taking large quantum numbers.
This should be true even for small $\lambda $, provided one takes
sufficiently large quantum
excitation numbers, such as the angular momentum.
{}For small $\lambda $, the supergravity background will receive
important corrections, but not in the region $u\gg u_0/\lambda $.



Among the different solutions of the previous sections, there are
several ones which can entirely lie in the $u\gg u_0/\lambda $ region.
These are the configurations: 1) \srsun , 2) \sunsqur \ and  3) \sali .
Increasing the frequency $\omega_1 $ around $\theta_1 $ produces the
effect
that the equilibrium position $u_1$ of the string moves towards larger
values of $u$. In the cases 1) and 2), demanding that  $u_1\gg
u_0/\lambda $
requires $\omega_1\gg n_1/\lambda^6$. For $\lambda <1$, 
this implies that $ p_{\theta_1}
\gg n_1/\lambda ^8$.
In this limit, the energy formulas in both cases 1) and 2) 
are
given by
\eqn\hhg{
1),\ 2):\ \ \ \ E_{\rm KK}\cong 3 u_0 p_{\theta_1} \gg 
{u_0 n_1\over \lambda^8}\ .
}
We recall that the solution 2) \sunsqur \ exists only for a discrete
set of $\lambda $ parametrized by the quantum numbers 
$J_\varphi, \ p_{\theta_1},\ n_1$.
For the case 3) \sali , the string lies in the   $u_1\gg u_0/\lambda $
region
provided $\omega_1\gg \sqrt{2m^2+\omega^2} /\lambda ^2$, i.e.
 $ p_{\theta_1}
\gg \sqrt{2m^2+\omega^2} /\lambda ^4$. This implies
 $ p_{\theta_1}
\gg m /\lambda ^4$. In this limit, the energy formula
becomes
\eqn\hhgg{
3)\ \ \ \ E_{\rm KK}
\cong 3 u_0 p_{\theta_1} \gg {u_0 m \over \lambda^4}\ .
}
For the reasons explained above, the energy formulas \hhg \
and \hhgg\ should
apply for small values of $\lambda $ provided $p_{\theta_1}$ is
sufficiently large as prescribed (to guarantee that  the string lies
in a small curvature region) and large angular momentum (to suppress
quantum sigma model corrections relative to the leading contribution
to the energy).

These formulas indicate a rapid decoupling of these Kaluza-Klein
states.

\newsec{ Description of supergravity glueballs from 
classical geodesic motion}


Masses of glueballs of the supergravity sector can be obtained by solving
the equations of motion corresponding to supergravity fluctuations 
\refs{\witten - \csakk }.
 In particular, the spectrum of a scalar glueball 
$0^{++}$ is obtained  from the supergravity equation for 
the dilaton mode fluctuation,
\eqn\lapl{
\partial_\mu(\sqrt{g}e^{-2\phi}g^{\mu\nu}\partial_\nu \tilde \phi) = 0\,.
}
Setting  
\eqn\fluctphi{\tilde \phi= e^{i k\cdot x} \chi(u)\,,
}
one obtains 
\eqn\laplbis{
{1\over u^3}\partial_u[u(u^6-u_0^6)\partial_u\chi(u)] + M^2
\chi(u)=0\,,\ \ \ \ M^2=-k^2\ .
}
 Using the WKB method, for large masses one can approximate the
 spectrum by \mina 
\eqn\ssugra{
M\cong {\pi \over \xi} \sqrt{n(n+2)}\ u_0\cong {\pi \over \xi}\, n \,
u_0\ ,
}
with $n$ integer and
\eqn\laxi{
\xi = u_0\int_{u_0}^\infty du \ {1 \over u^2\sqrt{h(u)}}=
{\sqrt{\pi} \, \Gamma({7\over 6}) \over \Gamma({2\over 3}) }
\,,
\ \ \ \ \ h(u)=1-{u_0^6\over u^6}\ .
}

\smallskip

Now we will show that this glueball spectrum can be alternatively 
obtained by 
considering a  point-like string configuration oscillating along the 
variable $u$ between the horizon and infinity.
To be specific, we
consider a configuration: $u=u(\tau )\ , \
t=t(\tau )\ $, describing a point-like string which
oscillates along a meridian $\theta_1=\theta_0 $ of the ``cigar"
described by the coordinates $u$ and $\theta_1$, from $u=\infty$
to $u=u_0$ and then back to $u=\infty$ following the ``other side"
of the meridian, $\theta_1=\theta_0+ \pi$. In the conformal gauge
the equations of motion set
$$
\dot t={c\over u^3} \ ,\qquad \ \ \dot u =\pm  {c\over u}\sqrt{h(u)}\,,
$$
which describe a null geodesic in the background \metric .
The plus sign corresponds to the branch describing the string coming from
$\infty $ to $u=u_0$, whereas the minus sign corresponds to the branch
 describing the string coming from $u=u_0$ to
$\infty $.
The (target-space) time elapsed in a motion from $u=\infty$ to $u_0$
and back
to $u=\infty$ is
\eqn\deltat{
\Delta T = \int dt = 2\int_{u_0}^\infty {\dot t\over \dot u} d u =
2\int_{u_0}^\infty du\ {1 \over u^2\sqrt{h(u)}} =
{2\sqrt{\pi}\,\Gamma({7\over 6}) \over u_0\,\Gamma({2\over 3}) }=
{2\xi\over u_0}\,,
}
which is finite, so the motion is periodical. 
One can define the ``angular" frequency 
${\omega_0}:= {2\pi\over \Delta T} = {\pi u_0\over \xi}$. 

In order to determine the quantum energy levels, it is more
convenient to work with the Nambu-Goto action. We will
consider a more general string configuration
$u=u(\tau )\,, \ \theta_1=m\sigma\,, $ and set
the gauge $\tau =t $. The case $m=0$ will be recovered at the
end. With this configuration we get the Lagrangian
\eqn\nhy{
{\cal L}=-{4 \over 9} m \lambda u_0 \sqrt{ {u^6\over
u_0^6} - 1 - {\dot u^2u^2\over u_0^6} }\ .
}
Following \minahan , we perform a change of variables in order to obtain
a Hamiltonian in a form suitable for the application of the 
WKB approximation. We introduce a new variable $\zeta$ 
$$
{d\zeta\over d u} = { u\over \sqrt{u^6-u_0^6}} \,.
$$
This defines $u=u(\zeta)$.
The Lagrangian \nhy\ takes the form
\eqn\newlagr{
{\cal L}= -{4 \over 9}  m \lambda u_0
\sqrt{\big( {u(\zeta)^6 \over u_0^6}   -1\big)(1-\dot\zeta^2)}\ .
}
The Hamiltonian is then given by
\eqn\bgt{
H=\sqrt{p^2_\zeta + ({4 \over 9} m \lambda u_0)^2 ({u(\zeta
)^6\over
  u_0^6}-1 ) }\, .
}
Note that ${1 \over 2} H^2$
describes a one dimensional system with potential $V= {1 \over
2}({4 \over 9} {m \lambda})^2 ({u^6\over u_0^6} - 1)u_0^2$. 
The potential is zero
at the horizon $u=u_0$ and diverges at $u=\infty$. We can
consider $u-u_0$ taking positive and negative values as well,
with a symmetric potential and taking into account only even wave
functions. In this case the Bohr-Sommerfeld formula gives the
quantization rule ($n\gg 1$)
$$
(2n+{1 \over 2})\pi \cong   \int d \zeta
\sqrt{E^2 - ({4 \over 9} m \lambda u_0)^2 ({u(\zeta )^6\over u_0^6} - 1)}= 
2 \int_{u_0}^{u_1} d u\,
u\,\sqrt{{E^2\over u^6-u_0^6} - ({4 \over 9}  
{m \lambda\over u_0^2})^2} \,,
$$
with 
$$
u_1={u_0\over \sqrt{\eta}}\ , \ \ \ \ 
\eta\equiv {1\over \big(1+ ({9 E \over 4\, m \lambda u_0})^2\big)^{1/3}}\,.$$ 
For $m=0$ we obtain 
$$
E \cong (n+{1\over 4}) \omega_0 \cong n\omega_0   = {\pi \over \xi}\, n\,
u_0\ .
$$
This exactly agrees with  \ssugra .
Thus one can think of supergravity glueballs as point-like
strings oscillating in the radial direction.
Note that the classical configuration does not distinguish parity,
i.e. it equally describes 
$0^{++}$ or $0^{-+}$ glueballs. In \refs{\mina ,\csakk}, 
the difference in
their energies is seen to be subleading in radial number $n$, so as
expected
in the
classical limit $n\to\infty $ their energies are the same.

In order to obtain the energy formula for $m\neq 0$, we write
\eqn\bohr{
(2n+{1 \over 2})\pi  \cong   {E\over u_0} 
 \sqrt{{\eta\over 1- \eta^3}} \int_\eta^1
d y \sqrt{{1-y^3\over y^3-\eta^3}}\,, \ \ \ \ y={u^2\over u_1^2}\ .
}
We can expand \bohr\ for small $\eta$, corresponding to 
$E\gg m\lambda  u_0$. This leads to the expansion 
\eqn\bohrm{(2n+{1 \over 2})\pi = {E\over u_0}{2\sqrt{\pi}\,\Gamma({7\over 6})
\over \Gamma({2\over 3}) } - {2^{2\over
3}\sqrt{\pi}\,\Gamma({5\over 6}) \over 3^{2\over 3}\Gamma({4\over
3}) } \big({E\over u_0}\big)^{2\over 3}m^{1\over 3}\lambda^{1\over 3}+\dots
}
i.e. 
\eqn\fnk{
E \cong u_0\big( 2.59 \,n +1.33\, 
m^{1\over 3}\lambda^{1\over 3} n^{2\over 3} +\dots\big)\ .
}
A different limit is $\lambda\gg 1 $ so that $\eta \sim 1$.
In this case one can check that the energy asymptotically
approaches the behavior $E\sim {\rm const. }\sqrt{\lambda }\ u_0$.
The derivative of the ratio ${E\over \sqrt{T_{\rm YM}}}$
with respect to $\lambda $ is negative for all $\lambda $.

\newsec{Concluding remarks}

We have determined the energy formula for different string
configurations in a string model dual to pure $SU(N)$ Yang-Mills in
3+1 dimensions.
The  dependence of the energy on $\lambda $ 
can be physically understood as follows.
The energy of a rotating string which is wound in some direction
has typically a kinetic contribution proportional to the
angular momentum, and a ``rest mass'' contribution proportional to the
string tension times the winding number times the length. 
For angular momentum in one
of the 
$S^4$ directions, the kinetic contribution is
typically independent of $\lambda $ and this is what produces the
behavior 
$E\sim u_0 $ or, equivalently, 
${E\over \sqrt{T_{\rm YM}}}\sim {1\over\sqrt{\lambda }}$ 
at $\lambda $ small enough so that 
the ``rest mass'' contribution is negligible ($\lambda $ can still be
$\lambda \gg 1 $, depending on the quantum numbers).
For a string wound around $S^4$, the string tension $T_{S^4}$
is proportional
to $\lambda $ (just as
the string tension $T_{\rm YM}$ of the  strings in $\Re^3 $).
 As a result, the ratio  ${E\over \sqrt{T_{\rm YM}}}$
is of order $\sqrt{\lambda }$ at large $\lambda $. This is the
reason of the large $\lambda $ behavior of \esqur \ and \kkali .

{}For low angular momentum, or for
sufficiently large string tension, the circular strings winding the sphere
 become unstable. This is expected, since the string can slide off the
 big circle of the sphere. But 
what is remarkable is that the instability
precisely
appears at the same point where the derivative of
 ${E_{\rm KK}\over \sqrt{T_{\rm YM}}}$ with respect to $\lambda $
 changes sign. This phenomenon arose in three independent cases.
{} Consequently,  for all classically stable strings with 
non-trivial Kaluza-Klein quantum numbers examined here, 
the ratio ${E_{\rm KK}\over
\sqrt{T_{\rm YM} }} $ always increases as $\lambda $ is decreased.
This behavior of the energy also holds for supergravity glueballs.
On the other hand, the energy of all glueballs associated with strings 
(with trivial Kaluza-Klein
quantum numbers) lying on the horizon  exhibits the behavior
behavior  ${E \over
\sqrt{T_{\rm YM} }} \sim {\rm const}$.
In sum, 
the general pattern that we find is that 
the energy of all classically stable states considered here  
does not increase
faster than $\sqrt{\lambda }$ at large $\lambda $.
It would be interesting to see if this is a property of all states
in the string spectrum.

The fact that for all Kaluza-Klein states the ratio
${E_{\rm KK}\over \sqrt{T_{\rm YM} }}$ increases as $\lambda $
is decreased, while for all  3+1 strings lying at the horizon this ratio
remains constant,
could be viewed  as an indication of the presence of two
distinct scales which may separate as $\lambda $ is gradually decreased.
However,  one should keep in mind that in general  these results 
cannot be extrapolated to
$\lambda \ll 1$ (except for states living sufficiently far from the horizon as
those of section 4). Also, the energy of supergravity glueballs
exhibits the same
behavior of KK states, 
i.e. ${E \over \sqrt{T_{\rm YM} }}$ increases as $\lambda $
is decreased, and they are not expected to decouple,
since the dilaton fluctuation is dual to the operator ${\rm Tr}\ F^2$.

In section 5, we have discussed how to describe supergravity
glueballs in terms classical configurations oscillating in the 
radial direction. 
It would be interesting
to similarly describe
 glueballs of higher spin ($>2$) in terms of strings oscillating
in the radial direction with  rotation in the angle $\beta $.

\bigskip

{\bf Ackowledgements}:
We would like to thank   D. Mateos for a useful discussion.
J.M.P. and J.R. acknowledge partial support by the European Commission 
RTN programme under 
contract HPNR-CT-2000-00131 and
by MCYT FPA 2001-3598 and CIRIT GC 2001SGR-00065.

\listrefs
\bye